\documentclass{article}
\usepackage{amssymb}

\usepackage{graphicx}
\usepackage{amsmath}

\newtheorem{theorem}{Theorem}

\newtheorem{corollary}[theorem]{Corollary}

\newtheorem{definition}[theorem]{Definition}

\newtheorem{lemma}[theorem]{Lemma}

\newtheorem{problem}[theorem]{Problem}

\newenvironment{proof}[1][Proof]{\textbf{#1.} }{\ \rule{0.5em}{0.5em}}
\input{tcilatex}

\begin{document}

\title{Range Theorems for Quantum Probability and Entanglement}
\author{I. Pitowsky \\
Department of Philosophy, The Hebrew University}
\maketitle

\begin{abstract}
We consider the set of all matrices of the form $p_{ij}=tr[W(E_{i}\otimes
F_{j})]$ where $E_{i}$, $F_{j}$ are projections on a Hilbert space $H$, and $%
W$ is some state on $H\otimes H$. We derive the basic properties of this
set, compare it with the classical range of probability, and note how its
properties may be related to a geometric measures of entanglement.
\end{abstract}

\section{Introduction}

Let $n$ be a natural number and consider the space of $(n+1)\times (n+1)$
real matrices, which we shall denote by $\Re _{n+1}$.The indices of a matrix
$(a_{ij})\in \Re _{n+1}$ have range $0\leq i,j\leq n$. My aim in this paper
is to investigate the subset of $\Re _{n+1}$which is given by the following:

\begin{definition}
$bell(n)$ is the set of all matrices $(p_{ij})\in \Re _{n+1}$ with the
following properties: $p_{00}=1$ and there exist a \ finite dimensional
Hilbert space $H$, projections $E_{1},E_{2},...,E_{n},F_{1}$, $%
F_{2},...F_{n} $ in $H$, and a statistical operator $W$ on the tensor
product $H\otimes H$ such that $p_{i0}=tr[W(E_{i}\otimes I)]$,$%
\;p_{0j}=tr[W(I\otimes F_{j})]$, $\;p_{ij}=tr[W(E_{i}\otimes F_{j})]$, for $%
i,j=1,2,...,n$. Here $I$ is the unit matrix on $H$. In case $W$ is pure we
shall say that $(p_{ij})$ has a pure state representation.
\end{definition}

Thus, $bell(n)$ is the range of probability values that states on a tensor
product assign to quantum events. Of particular interest are the probability
values assigned by entangled states which violate at least one Bell
inequality. We shall compare these probability values with the values in $%
c(n),$ the classical range:

\begin{definition}
$c(n)$ is the set of all matrices $(p_{ij})\in \Re _{n+1}$ with the
following properties: $p_{00}=1$ and there exists a probability space $%
(X,\Sigma ,\mu )$, events $A_{1},A_{2},...,A_{n}$, $B_{1},B_{2},...B_{n}\in
\Sigma $, such that $p_{i0}=\mu (A_{i})$, $p_{0j}=\mu (B_{j})$, $p_{ij}=\mu
(A_{i}\cap B_{j})$ for $i,j=1,2,...,n$ .
\end{definition}

The set $c(n)$ has been completely characterized \cite{1}\cite{2}. It is a
polytope (the closed convex hull of finitely many matrices) whose vertices
are the following: Let $(\varepsilon _{1},\varepsilon _{2},...,\varepsilon
_{n})$, $(\delta _{1},\delta _{2},...\delta _{n})\in \{0,1\}^{n}$ , be any
two arbitrary $n$-vectors of zeroes and ones, define a matrix $(u_{ij})$ by $%
u_{00}=1$, $u_{i0}=\varepsilon _{i}$, $u_{0j}=\delta _{j}$, $%
u_{ij}=\varepsilon _{i}\delta _{j}$ for $i,j=1,2,...,n$. Each such choice
defines a vertex of $c(n)$, altogether $2^{2n}$ vertices.

Every convex polytope in a linear space has a dual description, firstly in
terms of its vertices and secondly in terms of its \textit{facets}, linear
inequalities which describe the half spaces that bound it. In the case of
the correlation polytopes $c(n)$, the inequalities include the Bell
inequalities, Clauser Horne inequalities, and other inequalities that arise
in the study of entangled states. The investigation of these inequalities
began a long time ago \cite{3}\cite{4}. Deriving the description of a
polytope in terms of inequalities from a description in terms of vertices is
called \textit{the hull problem. }It is algorithmically solvable, but in
case of the correlation polytope the computational complexity is high \cite
{5}. For small $n$ the problem can be solved fairly quickly on a personal
computer \cite{6}\cite{7}. In the case $n=2$, the number of inequalities is $%
24$ and they include the Clauser Horne inequalities, for $n=3$ there are $%
684 $ inequalities!

In the next section I shall prove that $bell(n)$ is convex, and $%
bell(n)\supset c(n)$. A more detailed description is possible if we
concentrate on a special subset of $bell(n)$: Denote by $bell_{0}(n)$ the
subset of $\Re _{n+1}$ which is defined like $bell(n)$ but with the
additional conditions on the marginals $p_{i0}=p_{0j}=\frac{1}{2}$ for $%
i,j=1,2,...,n.$ Following \cite{8} we shall give in the third section a
simple characterization of the elements of $bell_{0}(n)$ . Of course, $%
bell_{0}(n)$ is also convex and $bell_{0}(n)\supset c_{0}(n)$ where $%
c_{0}(n) $ is the subset of $c(n)$ defined by the same conditions.

It is also interesting to compare $bell(n)$ with another quantum range. Note
that the tensor product plays the logical role of conjunction in quantum
mechanics. Thus measuring $E_{i}\otimes F_{j}$ consists of measuring $E_{i}$
``on the left'' \textbf{and}\textit{\ }$F_{j}$ ``on the right''. This in
analogy with the classical case where $A_{i}\cap B_{j}$ is the event $A_{i}$
\textbf{and}\textit{\ }$B_{j}$. However, the tensor product is not the most
general form of conjunction in quantum mechanics. Thus, Birkhoff and von
Neumann \cite{9} suggested that the quantum analogue of ``\textbf{and''}
should be subspace intersection. This leads to the following definition

\begin{definition}
$q(n)$ is the set of all matrices $(p_{ij})\in \Re _{n+1}$ with the
following properties: $p_{00}=1$ and there exist a Hilbert space $H$,
projections $E_{1},E_{2},...,E_{n}$, $F_{1},F_{2},...F_{n}$ in $H$ which do
not necessarily commute, and a statistical operator $W$ on $H$ such that $%
p_{i0}=tr(WE_{i})$,$\;p_{0j}=tr(WF_{j})$,$\;p_{ij}=tr[W(E_{i}\wedge F_{j})]$%
, for $i,j=1,2,...,n$. Here $E_{i}\wedge F_{j}$ is the projection on $%
E_{i}(H)\cap F_{j}(H)$.
\end{definition}

The set $q(n)$ has also been completely characterized\ \cite{1}\cite{2}. It
is convex but not (relatively) closed. Its closure in $\Re _{n+1}$ is a
polytope with vertices $(u_{ij})$ which are all the zero-one matrices
satisfying $u_{00}=1$, $u_{i0}\geq u_{ij}$, $u_{0j}\geq u_{ij}$ for $%
i,j=1,2,...,n$. We shall see that $q(n)\supset bell(n)\supset c(n)$.

Now, suppose that $W$ is a \textit{fixed} state on a tensor product $%
H\otimes H$. Let $n\geq 2$ be a natural number. \textit{The trajectory of }$%
W $\textit{\ in }$\Re _{n+1}$\textit{, }denoted by $b(W,n)$, is the set of
all matrices in $bell(n)$ which can be formed by applying $W$ to arbitrary $%
n $ projections on the ``left'' space and arbitrary $n$ projections on the
''right''. Assume $W$ is pure and $n\geq 2$. We shall see that if $W$ is a
product state then $b(W,n)\subset c(n)$, otherwise, $b(W,n)\nsubseteq c(n)$.
Hence, if $W$ is entangled then for all $n\geq 2$, parts of $b(W,n)$ lie
outside of $c(n)$. The maximal distance between $b(W,n)$ and $c(n)$ may
serve as a geometric measure of the entanglement of $W$.

Two remarks should be made at this point: (1) I have chosen the number of
projections on the ``left'' to be identical to the number on the ``right''.
There is no loss of generality in that, since the $n\times m$ case can be
imbedded in the $\max (m,n)\times \max (m,n)$ case by adding zero
projections. (2) many of the results that follow can be extended to
multipartite cases.

\section{The Set $bell(n)$}

\begin{theorem}
$bell(n)$ is convex and $q(n)\supset bell(n)\supset c(n)$.

\begin{proof}
Assume that $(p_{ij}),(q_{ij})\in bell(n)$ and let $0\leq \lambda \leq 1$,
we shall show that $(\lambda p_{ij}+(1-\lambda )q_{ij})\in bell(n)$. By
assumption there exist a finite dimensional Hilbert space $H$, projections $%
E_{0}=I,E_{1},E_{2},...,E_{n}$, and $F_{0}=I,F_{1},F_{2},...F_{n}$ in $H$,
and a statistical operator $W$ on the tensor product $H\otimes H$ such that $%
p_{ij}=tr[W(E_{i}\otimes F_{j})],$ for $i,j=0,1,2,...,n$. Similarly, there
exist a finite dimensional Hilbert space $H^{`}$, projections $%
E_{0}^{^{`}}=I^{`},E`_{1},E`_{2},...,E`_{n}$, and $%
F`_{0}=I`,F`_{1},F`_{2},...F`_{n}$ in $H`$, and a statistical operator $W`$
on the tensor product $H`\otimes H`$ such that $q_{ij}=tr[W`(E`_{i}\otimes
F`_{j})],$ for $i,j=0,1,2,...,n$. (Here $I`$ stands for the unit in $H`$).
Now, let $H``=H\oplus H`$ be the\textit{\ direct sum} of $H$ and $H`$, then $%
H``$ is finite dimensional. define $E``_{i}=E_{i}\oplus E`_{i}$ and $%
F``_{j}=F_{j}\oplus F`_{j}$. \ Since $W$ is a state in $H\otimes H$ it can
be represented as $W=\sum_{k}\lambda _{k}\left| \Phi _{k}\right\rangle
\left\langle \Phi _{k}\right| $ with $\lambda _{k}\geq 0,\sum_{k}\lambda
_{k}=1$, and $\left| \Phi _{k}\right\rangle $ unit vectors in $H$. Each
vector $\left| \Phi _{k}\right\rangle $ can be identified as a vector $%
\left| \Phi _{k}^{\ast }\right\rangle $ on $H``\otimes H``$ as follows: If $%
\left| \Phi _{k}\right\rangle =\sum_{l}c_{l}\left| \alpha _{l}\right\rangle
\left| \beta _{l}\right\rangle $ is the Schmidt decomposition of $\left|
\Phi _{k}\right\rangle $, we shall identify it with $\left| \Phi _{k}^{\ast
}\right\rangle =\sum_{l}c_{l}(\left| \alpha _{k}\right\rangle \oplus
0`)\otimes (\left| \beta _{l}\right\rangle \oplus 0`)\in $ $H``\otimes H``$,
where $0`$ stands for the zero vector of $H`$. Now identify $W$ as the state
$W$ $^{\ast }=\sum_{k}\lambda _{k}\left| \Phi _{k}^{\ast }\right\rangle
\left\langle \Phi _{k}^{\ast }\right| $ on $H``\otimes H``$.

If $W^{\prime }$ is similarly represented on $H`\otimes H`$ in terms of
vectors $\left| \Phi `_{k}\right\rangle =\sum_{k}c_{k}^{\prime }\left|
\alpha _{k}^{\prime }\right\rangle \left| \beta _{k}^{\prime }\right\rangle $%
, we can identify each $\left| \Phi `_{k}\right\rangle $ as a vector in $%
H``\otimes H``$, namely $\ \sum_{k}c`_{k}(0\oplus \left| \alpha
`_{k}\right\rangle )\otimes (0\oplus \left| \beta `_{k}\right\rangle )$. The
state $W^{^{\ast }\prime }$ is similarly identified as a state on $%
H``\otimes H``$. With this we can define the state $W``=\lambda W^{\ast
}+(1-\lambda )W^{\ast \prime }$ on $H``\otimes H``$. It is straightforward
to see that $tr[W``(E``_{i}\otimes F``_{j})]=\lambda p_{ij}+(1-\lambda
)q_{ij}$.

To prove that $bell(n)\supset c(n)$ let $(\varepsilon _{1},\varepsilon
_{2},...,\varepsilon _{n}),(\delta _{1},\delta _{2},...\delta _{n})\in
\{0,1\}^{n}$ be two zero-one vectors. Let $H$ be an arbitrary Hilbert space.
Define projections on $H$ by
\begin{equation*}
E_{0}=F_{0}=I\;\;\;\;\;E_{i}=\left\{
\begin{array}{c}
0\;\;\varepsilon _{i}=0 \\
I\;\;\ \varepsilon _{i}=1
\end{array}
\right. \;\;\;\;\;F_{j}=\left\{
\begin{array}{c}
0\;\;\delta _{j}=0 \\
I\;\;\delta _{j}=1
\end{array}
\right.
\end{equation*}
and let $W$ be any state. Then $tr[W(E_{i}\otimes I)]=\varepsilon _{i}$,$%
\;tr[W(I\otimes F_{j})]=\delta _{j}$,$\;tr[W(E_{i}\otimes
F_{j})]=\varepsilon _{i}\delta _{j}$. Hence, $bell(n)$ contains the vertices
of $c(n)$. Since $bell(n)$ is convex it contains all convex combinations of
the vertices. Therefore, $bell(n)\supset c(n)$. The other inclusion $%
q(n)\supset bell(n)$ is trivial, since $E_{i}\otimes F_{j}=(E_{i}\otimes
I)\wedge (I\otimes F_{j})$.
\end{proof}
\end{theorem}

We shall denote by $bell_{+}(n)$ the set of all matrices $(q_{ij})\in \Re
_{n+1}$ with the following property: there exist a \ finite dimensional
Hilbert space $H$, semi- definite operators $A_{0}=I,A_{1},A_{2},...,A_{n}$,
$B_{0}=I,B_{1},B_{2},...B_{n}$ in $H$, with $spectrum[A_{i}]\subset \lbrack
0,1]$, $spectrum[B_{j}]\subset \lbrack 0,1]$, and a statistical operator $W$
on the tensor product $H\otimes H$ such that $\;q_{ij}=tr[W(A_{i}\otimes
B_{j})]$, for $i,j=0,1,...,n$. Obviously, $bell(n)\subseteq bell_{+}(n)$. It
is easy to see, using identical technique to that of theorem 4, that $%
bell_{+}(n)$ is convex.

\begin{theorem}
$bell(n)=bell_{+}(n)$
\end{theorem}

\begin{proof}
Assume $q_{ij}=tr[W(A_{i}\otimes B_{j})]$. If $W$ is a mixture, $%
W=\sum_{k}\lambda _{i}\left| \Phi _{k}\right\rangle \left\langle \Phi
_{k}\right| $, $\lambda _{k}\geq 0$, $\sum_{k}\lambda _{k}=1$ then $%
q_{ij}=tr[W(A_{i}\otimes B_{j})]=\sum_{k}\lambda _{k}\left\langle \Phi
_{k}\right| A_{i}\otimes B_{j}\left| \Phi _{k}\right\rangle $ \ is a convex
combination of elements of $bell_{+}(n)$ that have a pure state
representation. Hence we can assume that $W$ is pure.

If \textbf{\ }$q_{ij}=\left\langle \Phi \right| A_{i}\otimes B_{j}\left|
\Phi \right\rangle $ and (at least) one of the $A_{i}$'s or $B_{j}$'s is not
a projection operator then $(q_{ij})$ is a convex combination. Suppose, for
example, that $A_{1}$ is not a projection operator. Then by the spectral
theorem we can write $A_{1}=$ $\sum_{k=1}^{l}\eta _{k}E^{k}$ with $1>\eta
_{1}>\eta _{2}>...>\eta _{l}>0$ and $E^{k}$ are pairwise orthogonal
projections, $E^{k}E^{r}=E^{r}E^{k}=0$. Hence for $j=0,1,...,n$
\begin{eqnarray*}
&&\;\;\;q_{1j}=\left\langle \Phi \right| A_{1}\otimes B_{j}\left| \Phi
\right\rangle =\eta _{l}\left\langle \Phi \right|
(E^{1}+E^{2}+...+E^{l})\otimes B_{j}\left| \Phi \right\rangle + \\
&&(\eta _{l-1}-\eta _{l})\left\langle \Phi \right|
(E^{1}+E^{2}+...+E^{l-1})\otimes B_{j}\left| \Phi \right\rangle +... \\
&&...+(\eta _{1}-\eta _{2})\left\langle \Phi \right| E^{1}\otimes
B_{j}\left| \Phi \right\rangle
\end{eqnarray*}
Note that $\eta _{l}+$ $(\eta _{l-1}-\eta _{l})+...+(\eta _{1}-\eta
_{2})=\eta _{1}\leq 1$, also $E^{1}+E^{2}+...+E^{k}$ are projection
operators. Now, for $k=1,2,...,l$ define
\begin{equation*}
A_{0}^{k}=B_{0}^{k}=I,\;\;\;\;\;A_{i}^{k}=\left\{
\begin{array}{c}
E^{1}+E^{2}+...+E^{k}\;\;i=1 \\
A_{i}\;\;\;\;\;\;\;\;\qquad \qquad \hspace{0.1cm}\ i>1
\end{array}
\right. ,\;\;\;\;\;B_{j}^{k}=B_{j}
\end{equation*}
Also put
\begin{equation*}
A_{0}^{l+1}=B_{0}^{l+1}=I,\;\;\;\;\;A_{i}^{l+1}=\left\{
\begin{array}{c}
0\qquad \;i=1 \\
A_{i}\qquad i>1
\end{array}
\right. ,\;\;\;\;\;B_{j}^{l+1}=B_{j}
\end{equation*}
then
\begin{eqnarray*}
q_{ij} &=&(\eta _{1}-\eta _{2})\left\langle \Phi \right| A_{1}^{1}\otimes
B_{j}^{1}\left| \Phi \right\rangle +....+(\eta _{l-1}-\eta _{l})\left\langle
\Phi \right| A_{1}^{l-1}\otimes B_{j}^{l-1}\left| \Phi \right\rangle + \\
&&\qquad +\eta _{l}\left\langle \Phi \right| A_{1}^{l}\otimes
B_{j}^{l}\left| \Phi \right\rangle +(1-\eta _{1})\left\langle \Phi \right|
A_{1}^{l+1}\otimes B_{j}^{l+1}\left| \Phi \right\rangle
\end{eqnarray*}
Combining the two stages we see that every element of $bell_{+}(n)$ -and
thus also of $bell(n)$- can be written as a convex combination of matrices
of the form $e_{ij}=\left\langle \Phi \right| E_{i}\otimes F_{j}\left| \Phi
\right\rangle $. Such matrices belong to $bell(n)$, hence, by convexity $%
bell(n)=bell_{+}(n)$. \ \ \
\end{proof}

Let $H$ be a Hilbert space of a finite dimension $m$. Let a unit vector $%
\left| \Phi \right\rangle $ in $H\otimes H$ be given in the Schmidt form $%
\left| \Phi \right\rangle =\sum_{i}c_{i}\left| \alpha _{i}\right\rangle
\left| \beta _{i}\right\rangle $ where $c_{i}$ are real and non-negative $%
\sum_{j}c_{i}^{2}=1$, and $\{\left| \alpha _{i}\right\rangle \}$, and $\
\{\left| \beta _{i}\right\rangle \}$, $\ i=1,2,...,m$ two orthonormal bases
in $H$. If $E,F$ are projections in $H$ we have for $W=\left| \Phi
\right\rangle \left\langle \Phi \right| $: $tr[W(E\otimes F)]=\left\langle
\Phi \right| E\otimes F\left| \Phi \right\rangle
=\sum\limits_{ij}c_{i}c_{j}\left\langle \alpha _{i}\right| E\left| \alpha
_{j}\right\rangle \left\langle \beta _{i}\right| F\left| \beta
_{j}\right\rangle $. Let $C$ be the diagonal matrix with $%
c_{1},c_{2},...,c_{m}$ on the diagonal, put $E_{ij}=\left\langle \alpha
_{i}\right| E\left| \alpha _{j}\right\rangle $, and $F_{ij}=\left\langle
\beta _{i}\right| F\left| \beta _{j}\right\rangle $ then $%
tr(CECF)=\sum_{ij}c_{i}c_{j}E_{ij}F_{ji}$. Now, define $F_{ij}^{\ast }=$ $%
F_{ji}=\left\langle \beta _{j}\right| F\left| \beta _{i}\right\rangle $ \
and note that $F^{\ast }$ \ is also a projection since $(F^{\ast
2})_{ik}=\sum_{j}F_{ij}^{\ast }F_{jk}^{\ast }$ $=\sum_{j}F_{ji}F_{kj}=%
\sum_{j}F_{kj}F_{ji}=(F^{2})_{ki}=F_{ki}=F_{ik}^{\ast }$. Hence we can write
$tr[W(E\otimes F)]=tr(CECF^{\ast })$. Combining this fact with theorem 2 we
have proved:

\begin{corollary}
: If \ $(p_{ij})\in bell(n)$ it can be represented as a convex combination
of matrices of the form $e_{ij}=tr(CE_{i}CF_{j})$ where $C$ is diagonal
positive and $tr(C^{2})=1$
\end{corollary}

For the sake of completeness we should say something about the closure (in
the Euclidean topology) of $bell(n)$, call it $\overline{bell}(n)$. If we
could find a natural number $N$, such that every $(p_{ij})\in bell(n)$ can
be represented on a Hilbert space of dimension $\leq N$, then $bell(n)=$ $%
\overline{bell}(n)$. Moreover, then the extreme points of $bell(n)$ must
have the form $e_{ij}=\left\langle \Phi _{s}\right| E_{i}\otimes F_{j}\left|
\Phi _{s}\right\rangle $. However, I was not able to prove that. (We shall
see below that in $bell_{0}(n)$ there is such a bound on the dimension).
What we can show, however, is that the elements of $\overline{bell}(n)$ have
a representation on a (possibly infinite dimesional) Hilbert space:

\begin{theorem}
If \ $(p_{ij})\in $ $\overline{bell}(n)$ then there exist a Hilbert space $H$%
, projections $E_{0}=I,E_{1},E_{2},...,E_{n}$, and $%
F_{0}=I,F_{1},F_{2},...F_{n}$ in $H$, and a statistical operator $W$ on the
tensor product $H\otimes H$ such that $p_{ij}=tr[W(E_{i}\otimes F_{j})],$
for $i,j=0,1,2,...,n$.

\begin{proof}
Let $\{(p_{ij}^{k})\}_{k=1,2,...}$ be a sequence of elements of $bell(n)$
which converges in the Euclidean topology to $(p_{ij})$. This means, in
particular, that $p_{ij}^{k}\rightarrow p_{ij}$ for all $i,j$ and therefore
also that $\underset{K\rightarrow \infty }{\lim }K^{-1}%
\sum_{n=1}^{k}p_{ij}^{k}=p_{ij}$. By assumption, for each $k$, there is a
finite dimensional Hilbert space $H_{k}$ \ projections $%
E_{0}^{k}=I,E_{1}^{k},E_{2}^{k},...,E_{n}^{k}$, and $%
F_{0}^{k}=I,F_{1}^{k},F_{2}^{k},...F_{n}^{k}$ in $H_{k}$ and a statistical
operator $W_{k}$ on the tensor product $H_{k}\otimes H_{k}$ such that $%
p_{ij}^{k}=tr[W(E_{i}^{k}\otimes F_{j}^{k})],$ for $i,j=0,1,2,...,n$.
Consider the space $\mathbb{H}_{K}=H_{1}\otimes H_{2}\otimes ...\otimes
H_{K} $ , the projections $\mathbb{E}_{i}^{K}=$ $E_{i}^{1}\otimes
E_{i}^{2}\otimes ...\otimes E_{i}^{K}$, \ $\mathbb{F}_{j}^{K}=$ $%
F_{j}^{1}\otimes F_{j}^{2}\otimes ...\otimes F_{j}^{K}$ and the state $%
\mathbb{W}_{K}$ on $\mathbb{H}_{K}\otimes \mathbb{H}_{K}$ defined as
follows: If $W_{r}=\sum_{kl}\lambda _{kl}\left| \alpha _{k}\right\rangle
\left\langle \alpha _{k}\right| \otimes \left| \beta _{l}\right\rangle
\left\langle \beta _{l}\right| $ \ put $W_{r}^{\ast }=\sum_{kl}\lambda _{kl}%
\underset{r}{(I\otimes ...\otimes \left| \alpha _{k}\right\rangle
\left\langle \alpha _{k}\right| \otimes ...\otimes I)}\otimes (I\otimes
...\otimes \underset{r}{\left| \beta _{l}\right\rangle \left\langle \beta
_{l}\right| }\otimes ...\otimes I)$ and $\mathbb{W}_{K}=K^{-1}(W_{1}^{\ast
}+W_{2}^{\ast }+...+W_{K}^{\ast })$. Then it is easy to see that $tr[\mathbb{%
W}_{K}(\mathbb{E}_{i}^{K}\otimes \mathbb{F}_{j}^{K})]=K^{-1}%
\sum_{n=1}^{k}p_{ij}^{k} $. Now , by a standard procedure \cite{10}\cite{11}
we can take the infinite tensor product limit $\mathbb{H}_{\infty }$ the
limits $\mathbb{E}_{i}^{\infty }$, and $\mathbb{F}_{j}^{\infty }$ and the
limit $\mathbb{W}_{\infty }$ on $\mathbb{H}_{\infty }\otimes \mathbb{H}%
_{\infty }$ with $tr[\mathbb{W}_{\infty }(\mathbb{E}_{i}^{\infty }\otimes
\mathbb{F}_{j}^{\infty })]=p_{ij}$. \
\end{proof}
\end{theorem}

\section{\protect\bigskip The set $bell_{0}(n)$}

Cirel'son (also spelled Tsirelson) \cite{8} considered the range of the
expectation values $s_{ij}=tr[W(A_{i}\otimes B_{j})]$ of operators $A_{i}$, $%
B_{j}$ which satisfy $spectrum[A_{i}]\subset \lbrack -1,1]$, $%
spectrum[B_{j}]\subset \lbrack -1,1]$. The $(s_{ij})$ is taken as an $%
n\times n$ matrix and we do not include the marginal values $tr[W(I\otimes
B_{j})]$ , and $tr[W(A_{i}\otimes I)]$ . This is a crucial point, as we
shall see later. Cirel'son's theorem is:

\begin{theorem}
The following conditions on an $n\times n$ matrix $(s_{ij})$ are equivalent:

a. There exists a Hilbert space $H$, Hermitian operators $%
A_{1,}A_{2},...A_{n}$, $B_{1,}B_{2},...B_{n}$, and a state $W$ on $H\otimes
H $ such that $spectrum[A_{i}]\subset \lbrack -1,1]$, $spectrum[B_{j}]%
\subset \lbrack -1,1]$ and $s_{ij}=tr[W(A_{i}\otimes B_{j})]$ for $%
i,j=1,2,...,n$.

b. The same as in 1, but with the additional conditions: $A_{i}^{2}=I$, $%
B_{j}^{2}=I$, $\ tr[W(A_{i}\otimes I)]=0$, $tr[W(I\otimes B_{j})]=0$, $\
A_{i_{1}}A_{i_{2}}+A_{i_{2}}A_{i_{1}}$ is proportional to $I$ \ for all $%
i_{1},i_{2}=1,2,...,n$, $\ B_{j_{1}}B_{j_{2}}+B_{j_{2}}B_{j_{1}}$ is
proportional to $I$ \ for all $j_{1},j_{2}=1,2,...,n$, and $\ \dim H\leq 2^{[%
\frac{n+1}{2}]}$.

c. There exist unit vectors $\mathbf{x}_{1,}\mathbf{x}_{2},...,\mathbf{x}%
_{n} $ and $\mathbf{y}_{1},\mathbf{y}_{2},...\mathbf{y}_{n}$ in the $2n$%
-dimensional real space $\mathbb{R}^{2n}$ such that $s_{ij}=\mathbf{x}%
_{i}\cdot \mathbf{y}_{j}$.
\end{theorem}

Call the set defined by the conditions of theorem 4 $tsirelson(n)$.To see
its connection with $bell_{0}(n)$ consider the second characterization in
theorem 4. If $A_{i}$ satisfies $A_{i}^{2}=I$ then by the spectral theorem
we can write $A_{i}=E_{i}-E_{i}^{\bot }$ where $E_{i}$ is a projection
operator, and $E_{i}^{\bot }$ is the projection on the subspace orthogonal
to $E_{i}(H)$. Similarly we can write $B_{j}=F_{j}-F_{j}^{\bot }$. \ Now,
from $tr[W(A_{i}\otimes I)]=0$ and the fact that $E_{i}+E_{i}^{\bot }=I$ we
conclude that $\ tr[W(E_{i}\otimes I)]=tr[W(E_{i}^{\bot }\otimes I)]=\frac{1%
}{2}$. Similarly, $tr[W(I\otimes F_{j})]=tr[W(I\otimes F_{j}^{\bot })]=\frac{%
1}{2}$. Denote $p_{ij}=tr[W(E_{i}\otimes F_{j})]$ then

\begin{equation*}
s_{ij}=tr[W(A_{i}\otimes B_{j})]=tr[W(E_{i}-E_{i}^{\bot })\otimes
(F_{j}-F_{j}^{\bot })]=4p_{ij}-1
\end{equation*}

Since $tr[W(E_{i}^{\bot }\otimes F_{j})]=\frac{1}{2}-p_{ij}$, $\
tr[W(E_{i}\otimes F_{j}^{\bot })]=\frac{1}{2}-p_{ij}$, \ $tr[W(E_{i}^{\bot
}\otimes F_{j}^{\bot })]=$ $p_{ij}$. Therefore, the map $s_{ij}\longmapsto
\frac{1}{4}(s_{ij}+1)$ maps $tsirelson(n)$ to $bell_{0}(n)$.

Conversely let $p_{ij}=tr[W(E_{i}\otimes F_{j})]$ be any element of $bell(n)$
(note! not necessarily $bell_{0}(n)$). Put $A_{i}=E_{i}-E_{i}^{\bot }$ and $%
B_{j}=F_{j}-F_{j}^{\bot }$, then by theorem 4a $s_{ij}=tr[W(A_{i}\otimes
B_{j})]\in tsirelson(n)$. Hence, the map $p_{ij}\longmapsto
4p_{ij}-2p_{i0}-2p_{0j}+1$ takes $bell(n)$ to $tsirelson(n)$ \ Combining the
two maps $bell(n)\longmapsto tsirelson(n)\longmapsto bell_{0}(n)$ we see
that $p_{i0}\longmapsto \frac{1}{2}$, $p_{0j}\longmapsto \frac{1}{2}$, $%
p_{ij}\longmapsto p_{ij}-\frac{1}{2}p_{i0}-\frac{1}{2}p_{0j}+\frac{1}{2}$
maps $bell(n)$ to $bell_{0}(n)$. Altogether we have shown

\begin{corollary}
(a) The set $bell_{0}(n)$ is convex and closed. If $\ (p_{ij})\in
bell_{0}(n) $ there is a Hilbert space $H$ with $\dim H\leq 2^{[\frac{n+1}{2}%
]}$, projections $E_{1},E_{2},...,E_{n},F_{1},F_{2},...F_{n}$ in $H$, and a
state $W$ such that $\ tr[W(E_{i}\otimes I)]=tr[W(I\otimes F_{j})]=\frac{1}{2%
}$ and $p_{ij}=tr[W(E_{i}\otimes F_{j})]$ for $i,j=1,2,...,n$. Moreover, we
can assume that $E_{i_{1}}E_{i_{2}}^{\bot }+E_{i_{2}}E_{i_{1}}^{\bot }$ and $%
F_{j_{1}}F_{j_{2}}^{\bot }+F_{j_{2}}F_{j_{1}}^{\bot }$ are proportional to $%
I $, for all $\ i_{1},i_{2},j_{1},j_{2}=1,2,..,n$.

(b) If $\ (p_{ij})\in bell_{0}(n)$ there exist unit vectors $\mathbf{x}_{1,}%
\mathbf{x}_{2},...,\mathbf{x}_{n}$ and $\mathbf{y}_{1},\mathbf{y}_{2},...%
\mathbf{y}_{n}$ in the $2n$-dimensional real space $\mathbb{R}^{2n}$ such
that $p_{ij}=\frac{1}{4}(\mathbf{x}_{i}\cdot \mathbf{y}_{j}+1)$. \ \
\end{corollary}

\section{Geometric Measures of Entanglement}

In recent years there have been numerous attempts to quantify the ''amount
of entanglement'' in a state defined on a tensor product of Hilbert spaces
\cite{12}. Most of these attempts are motivated by the concerns of quantum
information theory. Here I shall take a different route. Roughly, the
intuition is that the more entangled the state is the stronger the violation
of (at least one) Bell inequality. For simplicity I shall concentrate on
pure states.

\begin{definition}
Let $W=\left| \Psi \right\rangle \left\langle \Psi \right| $ be a fixed
state, $\left| \Psi \right\rangle =\sum_{i}c_{i}\left| \alpha
_{i}\right\rangle \left| \beta _{i}\right\rangle $ its Schmidt
decomposition. Then the trajectory of $W$ on $\Re _{n+1}$, denoted by $%
b(W,n) $, is the set of all matrices $(p_{ij})\in \Re _{n+1}$ that have the
form $p_{ij}=\left\langle \Psi \right| E_{i}\otimes F_{j}\left| \Psi
\right\rangle $, where $E_{0}=I,E_{1},E_{2},...,E_{n}$, and $%
F_{0}=I,F_{1},F_{2},...F_{n}$ \ are any projections in any finite
dimensional Hilbert space $H$ \ that contain $\{\left| \alpha
_{i}\right\rangle \}$, and $\ \{\left| \beta _{i}\right\rangle \}$.
\end{definition}

The connection between the trajectory and the classical range $c(n)$ is
given in the following.

\begin{lemma}
If $W=\left| \Psi \right\rangle \left\langle \Psi \right| $ is a product
state, $\left| \Psi \right\rangle =\left| \alpha \right\rangle \left| \beta
\right\rangle $ then $b(W,n)\subset c(n)$ for all $n$. Conversely, if $W$ is
not a product state then $b(W,n)\nsubseteq c(n)$ for all $n\geq 2$.
\end{lemma}

\begin{proof}
Suppose $\left| \Psi \right\rangle =\left| \alpha \right\rangle \left| \beta
\right\rangle $ and let $E_{0}=I,E_{1},E_{2},...,E_{n}$, and $%
E_{0}=I,E_{1},E_{2},...F_{n}$ be any projections in $H$. Consider the unit
square $[0,1]\times \lbrack 0,1]$ in the real plane $\mathbb{R}^{2}$ as a
probability space with $\Sigma $ the algebra of Borel subsets and $\mu $ the
uniform (Lebesgue) probability measure. Let $A_{i}$ be the subset of $%
[0,1]\times \lbrack 0,1]$ defined as $A_{i}=[0,\;\left\langle \alpha \right|
E_{i}\left| \alpha \right\rangle ]\times \lbrack 0,1]$ similarly define $%
B_{j}=[0,1]\times \lbrack 0,\;\left\langle \beta \right| F_{j}\left| \beta
\right\rangle ]$. Then $p_{i0}=\mu (A_{i})=\left\langle \alpha \right|
E_{i}\left| \alpha \right\rangle $, $p_{0j}=\mu (B_{j})=\left\langle \beta
\right| F_{j}\left| \beta \right\rangle $, $p_{ij}=\mu (A_{i}\cap
B_{j})=\left\langle \alpha \right| E_{i}\left| \alpha \right\rangle
\left\langle \beta \right| F_{j}\left| \beta \right\rangle $ for $%
i,j=1,2,...,n$ . Hence $p_{ij}=\left\langle \Psi \right| E_{i}\otimes
F_{j}\left| \Psi \right\rangle $ is an element of $c(n)$.

As for the converse, it follows from a theorem of Gisin and Peres\cite{13}.
They showed that if $\left| \Psi \right\rangle $ is not a product state then
one can choose projections $E_{0}=I,E_{1},E_{2}$ and $E_{0}=I,E_{1},E_{2}$
such that $p_{ij}=\left\langle \Psi \right| E_{i}\otimes F_{j}\left| \Psi
\right\rangle $ $ij=0,1,2$ violate the Clauser-Horne inequality. This
inequality is a facet inequality of $c(n)$ for all $n\geq 2$ \cite{2}.
Hence, $b(W,n)\nsubseteq c(n)$ for all $n\geq 2$. (It should be noted that
Gisin and Peres use observables with eigenvalues $\pm 1$. The transformation
to projection operators is the same as in the previous section).
\end{proof}

Let $\left\| \hspace{0.25cm}\right\| $ be a norn defined on $\Re _{n+1}$ ,
where $n\geq 2$ is fixed, and assume that $\ \left\| \hspace{0.25cm}\right\|
$ is continuous with respect to the Euclidean topology on $\Re _{n+1}$. Let $%
W=\left| \Psi \right\rangle \left\langle \Psi \right| $ be a pure state on $%
H\otimes H$, we shall define the entanglement measure associated with $%
\left\| \hspace{0.25cm}\right\| $ to be
\begin{equation}
\mathcal{E}(W)=\underset{(p_{ij})\in b(W,n)}{\sup }\ \underset{(q_{ij})\in
c(n)}{\min }\left\| (p_{ij})-(q_{ij})\right\|  \tag{1}
\end{equation}
The minimum in (1) is obtained for each $(p_{ij})\in b(W,n)$, because $%
\left\| \hspace{0.25cm}\right\| $ is continuous and $c(n)$ compact in the
Euclidean topology. From lemma it follows that $\mathcal{E}(W)=0$ if, and
only if $W$ is a product state.

\begin{problem}
Let $\left| \Psi \right\rangle =\sum_{i=1}^{m}c_{i}\left| \alpha
_{i}\right\rangle \left| \beta _{i}\right\rangle $ and $\left| \Phi
\right\rangle =\sum_{i=1}^{m}d_{i}\left| \delta _{i}\right\rangle \left|
\gamma _{i}\right\rangle $ be the Schmidt decompositions of two unit vectors
on $H\otimes H$. Assume $c_{1}\geq c_{2}\geq ...\geq c_{m}\geq 0$, and $%
d_{1}\geq d_{2}\geq ...\geq d_{m}\geq 0$. Recall that $\left| \Psi
\right\rangle $ \textit{majorizes }$\left| \Phi \right\rangle $ if $%
\sum_{i=1}^{k}c_{i}^{2}\geq \sum_{i=1}^{k}d_{i}^{2}$ for all $1\leq k\leq m$%
, in this case we shall denote $\left| \Psi \right\rangle $ $\succcurlyeq $%
\textit{\ }$\left| \Phi \right\rangle $. Under what conditions $\mathcal{E}%
(W)$ is monotone decreasing: $\left| \Psi \right\rangle $ $\succcurlyeq $%
\textit{\ }$\left| \Phi \right\rangle $ entails $\mathcal{E}(\left| \Psi
\right\rangle \left\langle \Psi \right| )\leq \mathcal{E}(\left| \Phi
\right\rangle \left\langle \Phi \right| )$
\end{problem}

Here the theorem of Nielsen may be helpful\cite{12}\cite{14}.

\begin{problem}
Does any of the familiar entanglement measures, in particular von Neumann's
entropy, have a geometric origin as above?
\end{problem}

I do not know the answer. A possible way to go is to use the uniqueness
theorems \cite{12}\cite{15}, and try determine if there is a geometric
measure which conforms with its conditions.

{\large Acknowledgment \ }This research is supported by the Israel Science
Foundation (grant number 787/99-01)

\bigskip

\bigskip

\end{document}